\documentclass[a4paper,11pt]{article}
\pdfoutput=1 

\usepackage{jinstpub} 

\title{\boldmath Characterization of the HZC Photonics XP82B20D and XP1805D Photomultiplier Tubes for Low-Temperature Applications}

\author[1]{D.\,van Eijk\note{Corresponding author.},}
\author{J.\,Dorant,}
\author{C.\,Wendt,}
\author{and A.\,Karle}

\affiliation{WIPAC, UW Madison, 222 West Washington Ave, Suite 500, Madison, WI 53703, USA}

\emailAdd{daan.vaneijk@icecube.wisc.edu}

\abstract{The IceCube Collaboration is investigating various types and manufacturers of photomultiplier tubes (PMT) for possible use in future optical modules. This report presents characterization results for two different types of HZC Photonics PMTs: the 3.5 inch XP82B20D and the 9 inch XP1805D. The results are in good agreement with the specifications as provided by the manufacturer. In addition, excellent noise behaviour is observed at the low temperatures relevant for possible use in IceCube optical modules.}


\begin{document}
\maketitle
\flushbottom

\section{Introduction}\label{sec:introduction}
Traditionally, photomultiplier tubes (PMT) with a relatively large diameter of 10 inch are used in large-volume neutrino experiments such as ANTARES \cite{Antares},  IceCube \cite{IceCube} and Baikal-GVD \cite{GVD}. More recently, the KM3NeT experiment \cite{KM3NeT} has adopted a strategy of implementing smaller 3 inch PMTs in multi-PMT optical modules. The IceCube Collaboration also plans to adopt this multi-PMT approach in future optical modules, both in the context of the IceCube Upgrade \cite{IceCubeUpgrade} and IceCube Gen2 \cite{IceCubeGen2}. To that end, various types of PMTs are currently under investigation. This report summarizes characterization results for two HZC Photonics \cite{hzcphotonics} PMTs: the 3.5 inch XP82B20D and the 9 inch XP1805D. Given the low temperatures down to -45\,$^{\circ}$C in the deep Antarctic ice where the IceCube experiment is located, special attention is given to low-temperature operation of the PMTs.

Section\,\ref{sec:setup} describes the experimental setup used for the characterization of the PMTs. In section\,\ref{sec:pulsechar}, methodologies to measure various PMT pulse characteristics and subsequent results are presented. Section\,\ref{sec:darknoise} focuses on dark noise characteristics with an emphasis on low-temperature behavior. 

\section{Experimental Setup}\label{sec:setup}

\subsection{PMT Types} 
\begin{figure}[ht]
    \centering
        \includegraphics[width=0.9\textwidth]{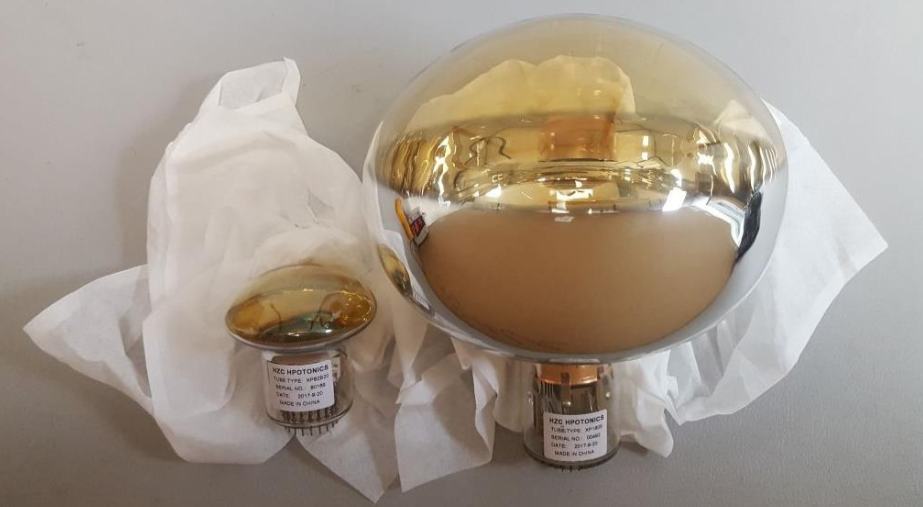}
        \caption{Picture of the two PMT types provided by HZC: a 3.5 inch XP82B20D PMT on the left and a 9 inch XP1805D PMT on the right.}
        \label{fig:pmts}
\end{figure}

\begin{figure}[ht]
    \centering
        \includegraphics[width=0.9\textwidth]{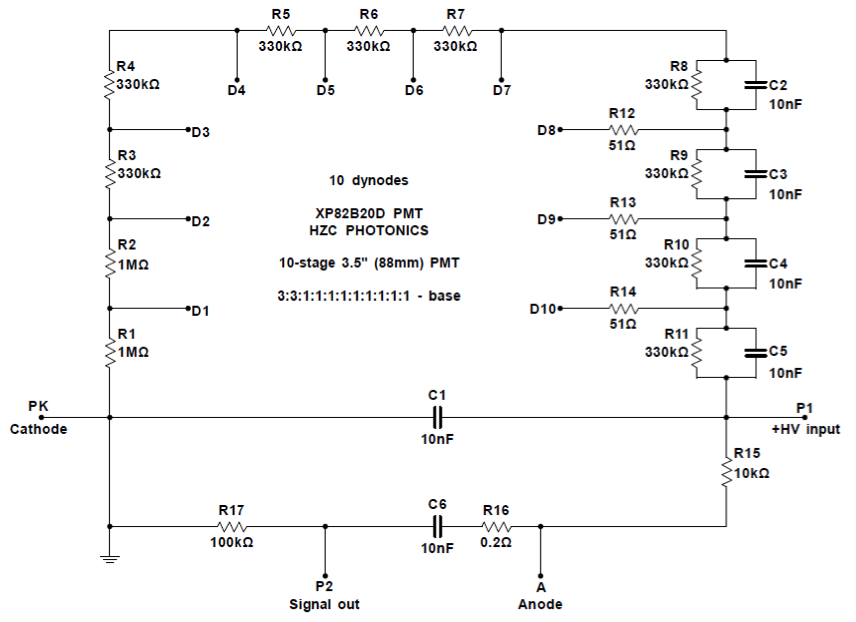}
        \caption{Circuit diagram of the HZC base with a voltage division of 3:3:1:1... for the XP82B20D PMTs.}
        \label{fig:base_circuit_diagram}
\end{figure}

Figure~\ref{fig:pmts} shows the two different HZC Photonics (referred to as HZC onward) PMT types that have been characterised in this study:
\begin{itemize}
\item XP82B20D (three pieces): 3.5 inch, 10-stage, high QE (28\% at 404 nm), borosilicate low K window, bi-alkali photocathode;
\item XP1805D (two pieces): 9 inch, 8-stage, borosilicate window, bi-alkali photocathode.
\end{itemize}
The PMTs were operated using positive polarity high voltage (HV) bases provided by HZC. As an example, the circuit diagram for the XP82B20D base with a voltage division of 3:3:1:1... is shown in figure~\ref{fig:base_circuit_diagram}.

\subsection{PMT Readout}
The PMTs are operated in pulse mode, using an oscilloscope\footnote{LeCroy Waverunner LT374L, bandwidth 500\,MHz} for visualization and analysis of the PMT pulse characteristics. The PMT base signal output (capacitively coupled to the anode) is connected to the oscilloscope without amplification using a coaxial cable and 50\,$\Omega$ termination. 

For dark noise characteristics, the PMT signals are read out using a custom-made PCB called microDAQ, which was developed both in the context of the IceTop Scintillator Upgrade \cite{IceTopUpgrade} and the CHIPS experiment \cite{CHIPS}. The PMT signal is amplified and then shaped by a passive component network, designed to make the ADC input voltage proportional to the pulse charge and nearly constant in time when sampled about 65\,ns after the discriminator trigger. In addition, the microDAQ adds a time stamp to all the PMT hits. 

\subsection{Dark Box and Freezer}
The PMT, its attached HV base and the microDAQ board are placed in a dark box that contains an optical fiber guiding photons from a laser pulser\footnote{Hamamatsu PLP-10, 405\,nm diode laser, 70\,ps FWHM pulses} to the center of the PMT photocathode area. The oscilloscope is triggered on a synchronization signal from the laser, which is set up to deliver low brightness pulses such that the PMT detects zero or sometimes one photon per pulse.

High voltage is provided to the base by an external analog HV supply\footnote{Power Designs 1570A} with an accuracy of 0.25\%. 

For temperature-dependent noise studies, the PMT stack is placed in a temperature-controlled dark freezer. Low-temperature PMT noise studies are of particular importance for possible use in future IceCube modules because of the ambient temperatures down to -45\,$^{\circ}$C in the deep Antarctic ice where the PMTs must ultimately operate.

\section{PMT Pulse Characteristics}\label{sec:pulsechar}

\subsection{Pulse Shape}
\begin{figure}[ht]
        \includegraphics[width=0.9\textwidth]{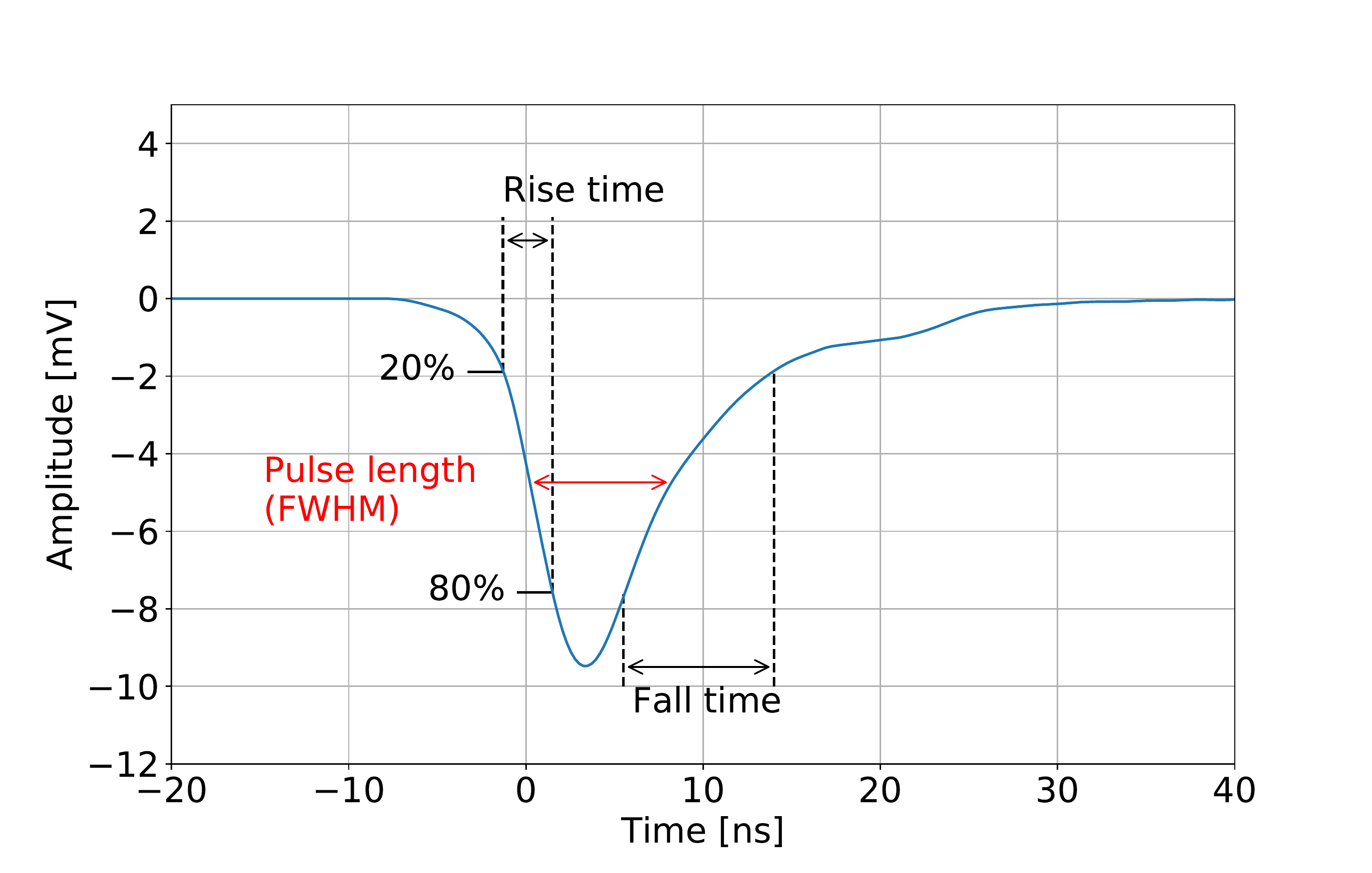}
        \caption{Typical pulse shape (averaged over 1000 pulses) for the 3.5 inch XP82B20D PMT with serial number SN80187 and the definition of pulse length at full-width half-maximum and rise and fall time at 20\% and 80\%. The PMT was operated at a gain of $1 \times 10^{7}$ (for an explanation of gain, see section \ref{sec:gain}).}
    	\label{fig:pulseshape}
\end{figure}

Figure~\ref{fig:pulseshape} shows an average pulse shape for one of the 3.5 inch XP82B20D PMTs and an explanation of various pulse shape characteristics. The rise (fall) time is the time taken by the signal to rise (fall) from 20\% (80\%) to 80\% (20\%) of the (negative) pulse amplitude. The pulse length is defined as the full-width of the pulse shape at half-maximum (FWHM). 

\subsection{Charge Distribution}
\begin{figure}[ht]
        \includegraphics[width=0.97\textwidth]{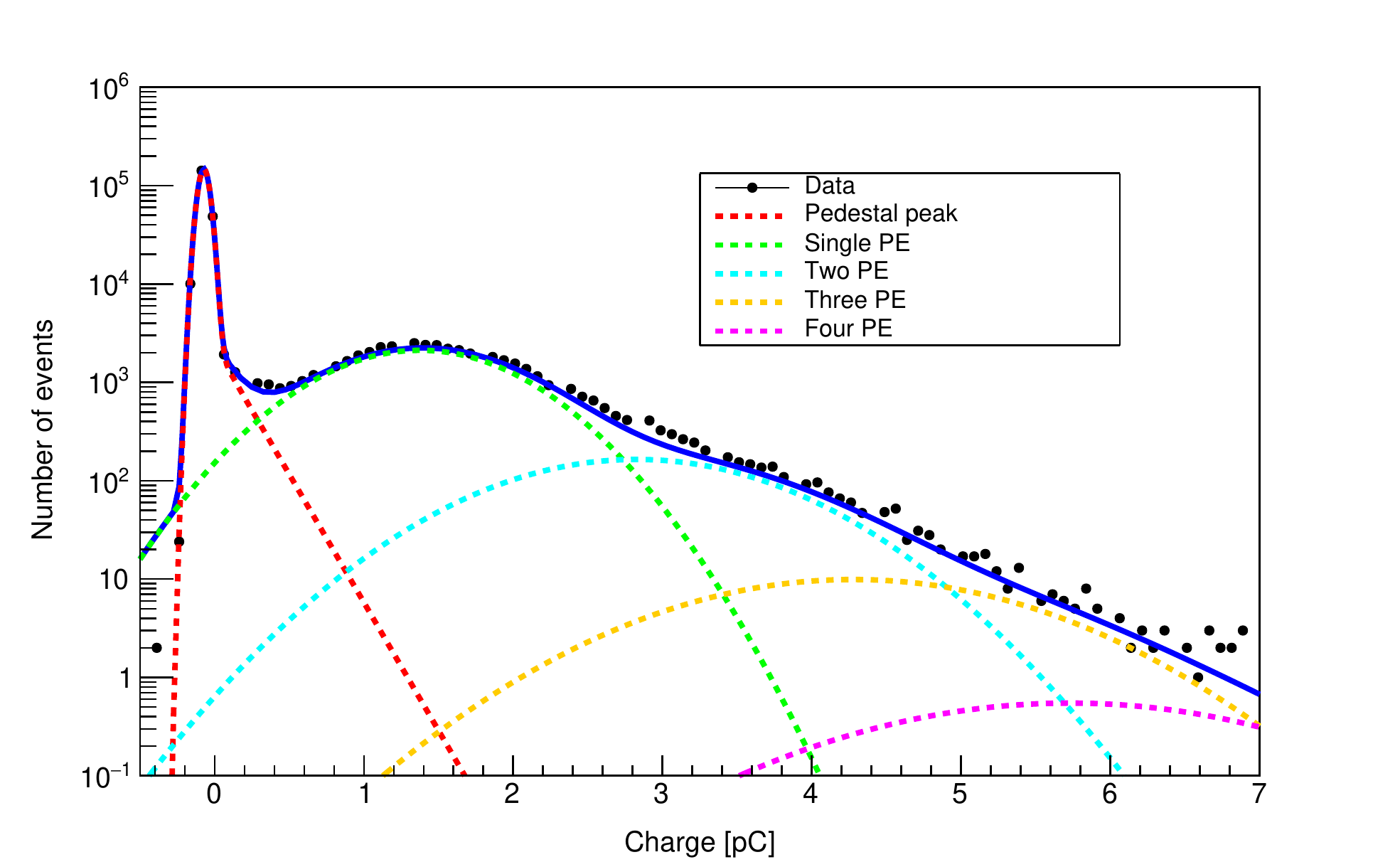}	
        \caption{The charge distribution is a histogram of time-integrated PMT pulse shapes divided by the oscilloscope's input impedance, hence every entry in the histogram is equal to the collected charge in a pulse. The fitted model is shown as a solid blue line. The various model components (the pedestal peak, a single photoelectron (PE) and multiple photoelectrons) are shown as dotted lines in different colors. Measurements were taken at room temperature.}
    	\label{fig:spespectrum}
\end{figure}

The time integral of the PMT pulse as measured by the oscilloscope divided by the oscilloscope's input impedance (50\,$\Omega$) equals the charge collected by the PMT. By making a histogram of these time-integrated pulses, one obtains the charge distribution that can be modelled according to \cite{PE_fit}. The model has various contributions, the largest being the pedestal peak corresponding to zero photoelectrons being present. At higher collected charge, the single and multi-photoelectron components start to contribute to the charge distribution as well. An example of a charge distribution from one of the 3.5 inch PMTs is shown in figure~\ref{fig:spespectrum}, including the fitted model.

\subsection{Gain Calibration}\label{sec:gain}
The PMT gain is defined as the number of detected electrons at the anode as a result of the amplification in the dynode structure of a single photoelectron (SPE) that hits the first dynode. This number corresponds to the distance between the SPE peak at charge $\mu_{\mathrm{SPE}}$ and the pedestal peak at charge $\mu_{\mathrm{pedestal}}$ in the charge distribution divided by the electron charge $Q_{\mathrm{e}}$:

\begin{equation}
G = \frac{\mu_{\mathrm{SPE}}-\mu_{\mathrm{pedestal}}}{ Q_\mathrm{e}} \quad .
\end{equation}

\begin{figure}[ht]
    \centering
        \includegraphics[width=0.9\textwidth]{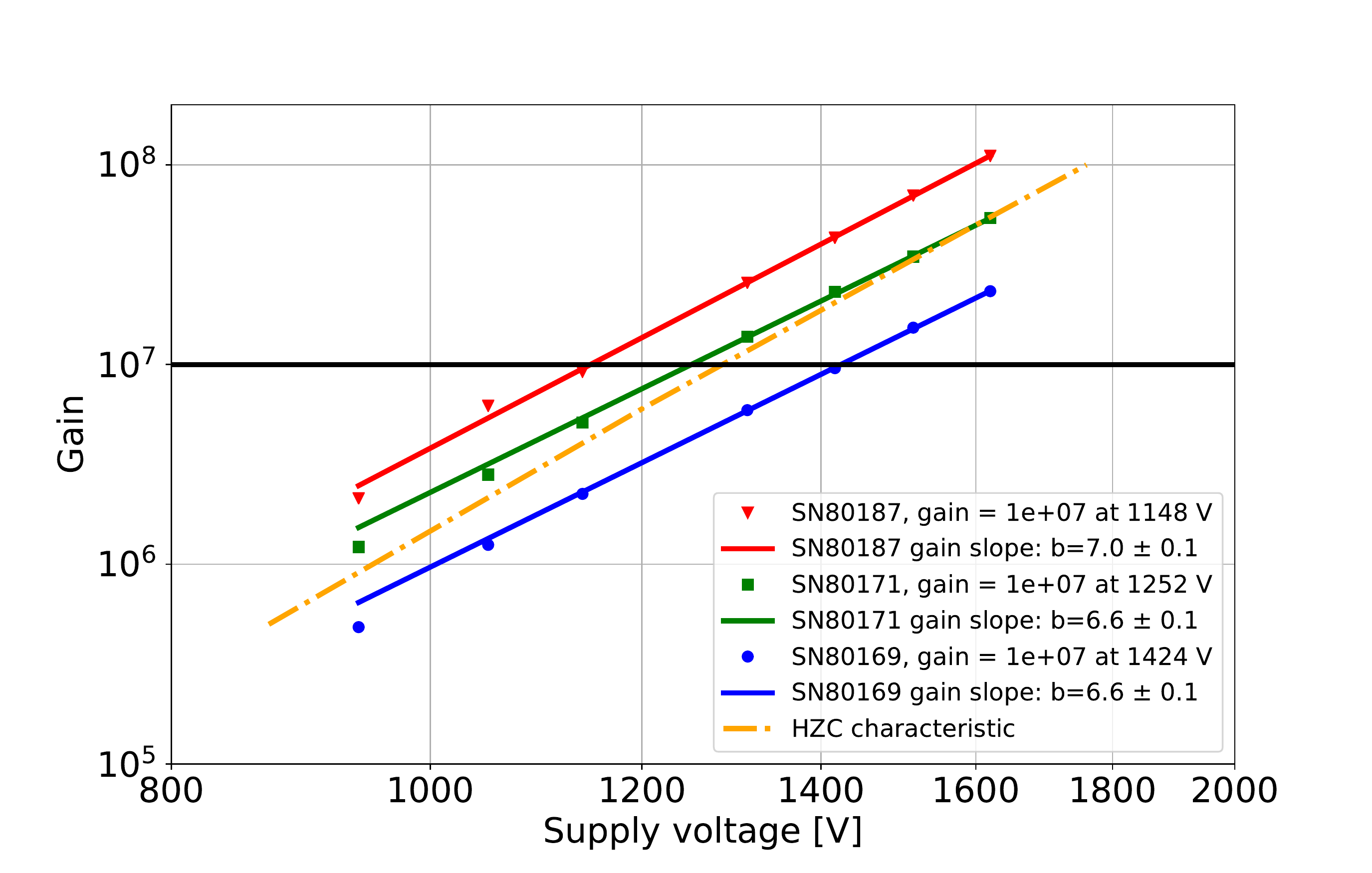}
        \caption{Gain calibration curves for the three 3.5 inch XP82B20D PMTs. The error bars for the supply voltage and gain data points are too small to be visible. The typical characteristic for this PMT type as provided by HZC is shown as the yellow dotted line and a gain of $1 \times 10^7$ is shown as a thick horizontal line. Measurements were taken at room temperature.}
        \label{fig:gainvshvsmall}
\end{figure}

By calculating the gain as a function of varying supply voltage one obtains a gain calibration curve that is fitted with a power law $a \times V^{b}$, where $V$ is the HV value, $a$ is a normalisation parameter and $b$ is the so-called gain slope as this parameter represents the slope of a gain calibration curve on a log-log plot. As an example, the gain calibration curves for the three 3.5 inch XP82B20D PMTs are shown in figure~\ref{fig:gainvshvsmall}.

Using the gain calibration curves, the HV value for a gain of $1 \times 10^{7}$ is determined, which corresponds to the typical operational gain in IceCube. These HV values are reported in table~\ref{tab:gaincalibration} and will be used to operate the PMTs for all following studies. Since HZC tested the PMTs at a gain of $3 \times 10^{6}$, this gain value is added to the table for a fair comparison. The discrepancies between the HV values at a gain of $3 \times 10^{6}$ as measured in this study and by HZC are likely due to different operating conditions. In particular, it is unclear whether the same type of HV bases were used for testing by the manufacturer.

\begin{table}[ht]
\begin{centering}
\footnotesize
\begin{tabular}{|lp{3.5cm}p{3.9cm}p{3.9cm}|}
\hline
\textbf{PMT} & \textbf{HV for gain $3 \times 10^{6}$ (HZC)} & \textbf{HV for gain $3 \times 10^{6}$ (this work)} & \textbf{HV for gain $1 \times 10^{7}$ (this work)} \\
\hline
SN80187 & 940 & 966 & 1148 \\
SN80171 & 1030 & 1042 & 1252 \\
SN80169 & 1160 & 1186 & 1424 \\
\hline
SN00386 & 1834 & 1731 & 2196\\
SN00492 & 1916 & 1713 & 2165\\
\hline
\end{tabular}
\caption{Gain calibration for all PMTs. The second column contains the HV values as measured by HZC at a gain of $3 \times 10^{6}$. The third and fourth columns report the HV values determined in this study for a gain of $3 \times 10^{6}$ and $1 \times 10^{7}$, respectively. The first three rows are the 3.5 inch XP82B20D, whereas the last two rows are the 9 inch XP1805D.}
\label{tab:gaincalibration}
\end{centering}
\end{table}

For completeness, the fitted slopes of the gain calibration curves for all tested PMTs are presented in summary table~\ref{tab:summarysmall} and summary table~\ref{tab:summarylarge} for the 3.5 inch XP82B20D PMTs and the 9 inch XP1805D PMTs, respectively. 

\begin{table}[ht]
\begin{centering}
\footnotesize
\begin{tabular}{|lp{1.8cm}p{2.0cm}p{2.0cm}p{2.2cm}|}
\hline
\textbf{PMT} & \textbf{FWHM [ns]} & \textbf{Rise time [ns]} & \textbf{Fall time [ns]} & \textbf{Peak-to-valley}\\
\hline
SN80169 & 7.0 & 2.6 & 7.2 & 3.6\\
SN80171 & 6.5 & 2.5 & 6.5 & 4.3\\
SN80187 & 8.2 & 2.8 & 8.5 & 3.1\\
\hline
\end{tabular}
\caption{Various pulse characteristics for averaged pulses measured at a gain of $1 \times 10^{7}$ for the 3.5 inch XP82B20D PMTs.}
\label{tab:pulsecharacteristics}
\end{centering}
\end{table}

Table~\ref{tab:pulsecharacteristics} reports the various pulse characteristics discussed in the previous sections for the 3.5 inch XP82B20D PMTs operating at a gain of $1 \times 10^{7}$. The peak-to-valley ratio is defined as the number of events at the mean of the SPE peak divided by the number of events at the minimum between the pedestal peak and the SPE peak.

\subsection{Transit Time and Early and Late Pulses}
The PMT transit time is the time between photoelectron emission and the moment when the PMT signal reaches its maximum amplitude. The relative transit time is defined here as the time difference between the external laser pulse that triggers the oscilloscope and the moment when the PMT signal reaches its maximum amplitude.

\begin{figure}[ht]
	\centering
	\includegraphics[width=0.9\textwidth]{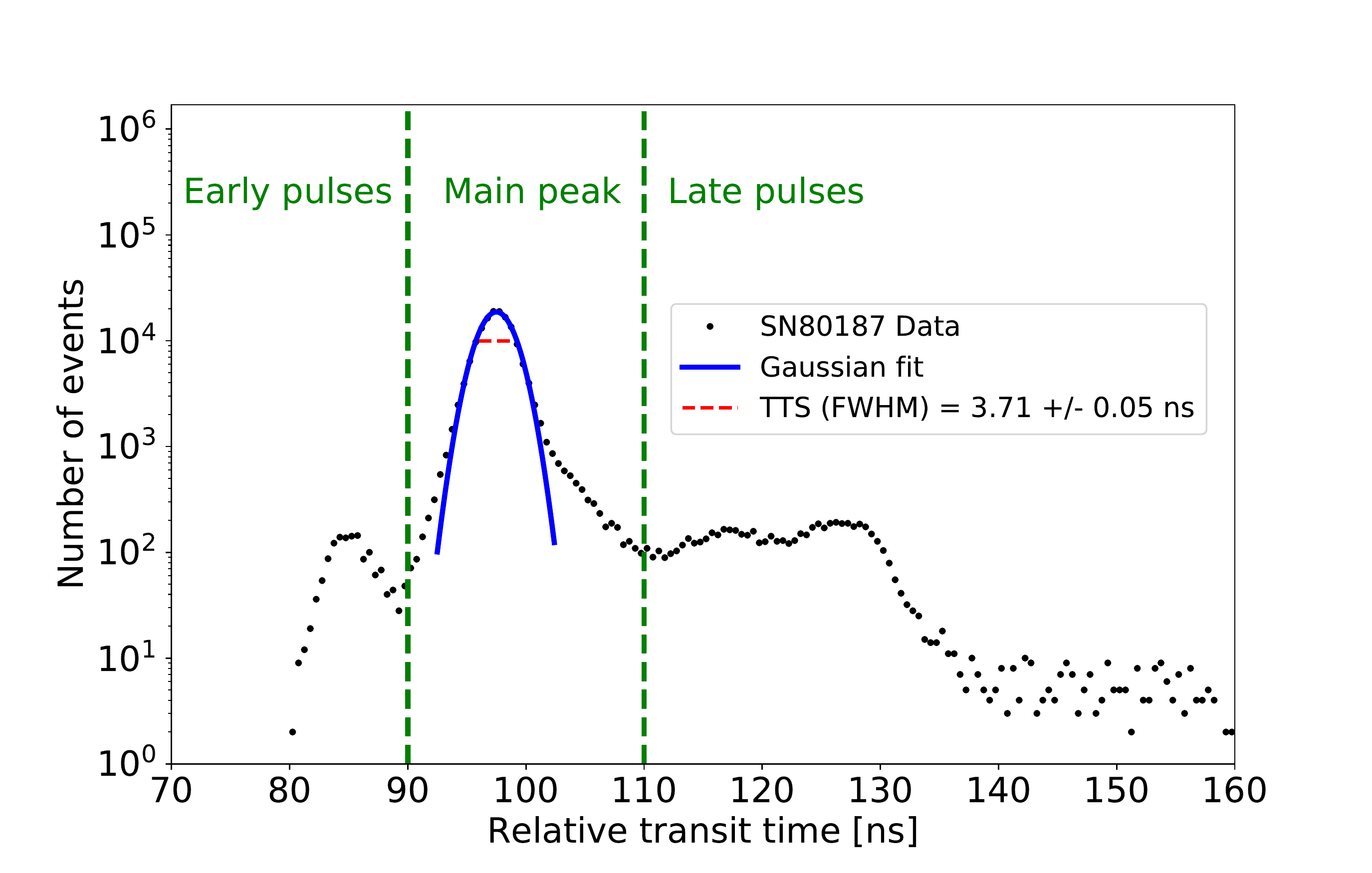}
	\caption{Relative transit time histogram for the 3.5 inch XP82B20D PMT with serial number SN80187. The main peak is located between the two vertical green lines. Early pulses and late pulses are located left and right of the main peak, respectively. The Gaussian fit to the main peak is shown in blue and the resulting transit time spread (TTS) value is indicated by the horizontal red dashed line. Measurements were taken at room temperature.}
\label{fig:pulsearrivalexample}
\end{figure}

In order to study the spread in PMT transit time, the relative transit time is binned. An example of the resulting relative transit time histogram is shown in figure~\ref{fig:pulsearrivalexample}. The histogram shows the main peak between the green vertical lines and spurious pulses arriving before (early pulses) and after (late pulses) the main peak. The number of early pulses and late pulses is calculated as a fraction of the total number of events. These fractions are summarised in table~\ref{tab:summarysmall} and table~\ref{tab:summarylarge} for the 3.5 inch XP82B20D PMTs and the 9 inch XP1805D PMTs, respectively.

\subsection{Transit Time Spread}
The transit time spread is defined as the FWHM\footnote{For a Gaussian distribution: $\mathrm{FWHM} = 2\sqrt{2\ln{2}}\,\sigma \approx 2.4\sigma$} of a Gaussian fit to the main peak of the relative transit time histogram, see figure~\ref{fig:pulsearrivalexample}. The transit time spread results are summarised in table~\ref{tab:summarysmall} and table~\ref{tab:summarylarge} for the 3.5 inch XP82B20D PMTs and the 9 inch XP1805D PMTs, respectively.

\section{Dark Noise Characteristics}\label{sec:darknoise}
For all dark noise studies, the PMTs are operated at the HV values for a gain of $1 \times 10^7$ at room temperature that were presented in section\,\ref{sec:gain}. To correct for gain changes, at each lower temperature the discriminator level of the microDAQ is adjusted to a value corresponding to 0.25 photoelectrons in the charge distribution. The microDAQ readout board operates in the so-called coarse time stamping mode, in which hit times are specified in bins of 5.6\,ns and each hit is followed by a dead time of 700\,ns.

\subsection{Total Dark Noise Rate}
The total dark noise rate is defined as the total number of pulses divided by the run time when the PMT is not illuminated. By binning the time difference between subsequent PMT pulses, various contributions to the total dark noise can be distinguished. 

\subsection{Correlated and Uncorrelated Noise}
\begin{figure}[ht]
    \centering
        \includegraphics[width=0.9\textwidth]{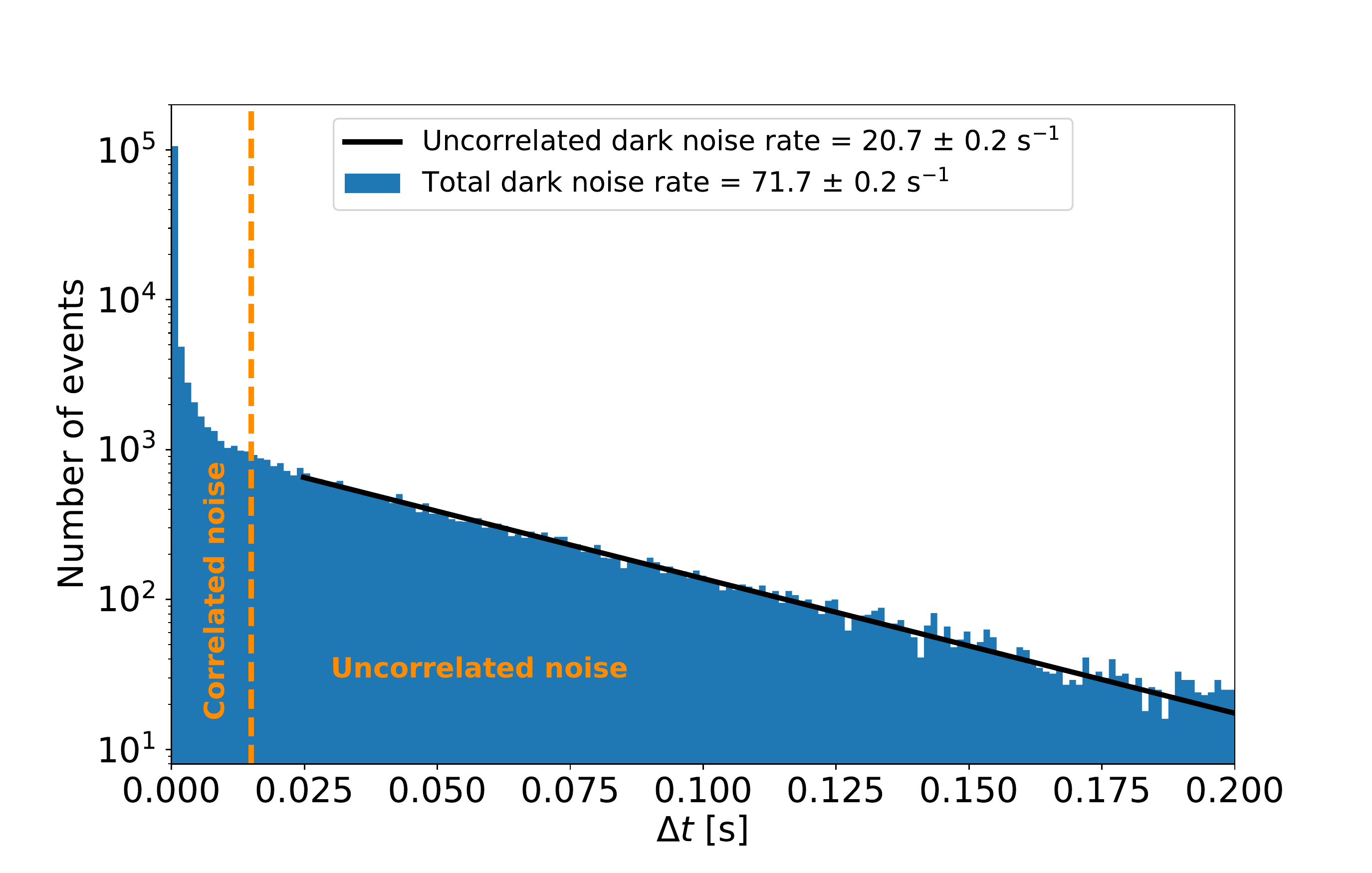}
        \caption{Histogram of the time difference between subsequent dark noise pulses for the 3.5 inch XP82B20D PMT with serial number SN80171 at a temperature of -30\,$^{\circ}$C. There are two main contributions to the total dark noise: the uncorrelated noise at large time differences and the correlated noise at small time differences.}
        \label{fig:examplerate}
\end{figure}

A typical histogram of the time differences between PMT pulses is shown in figure~\ref{fig:examplerate}. As expected, there is an exponentially falling contribution called the uncorrelated dark noise. However, at small time differences, there is an additional (non-exponential) contribution to the total dark noise rate. This contribution is called the correlated dark noise and could come from radioactive decays and scintillation in the PMT glass \cite{Lew, Meyer}. Both the total dark noise rate (total number of events divided by run time) and the uncorrelated dark noise rate (the larger time-difference events fitted with an exponential function) are reported here.

\subsection{Artificial Dead Time}
\begin{figure}[ht]
    \centering
        \includegraphics[width=0.9\textwidth]{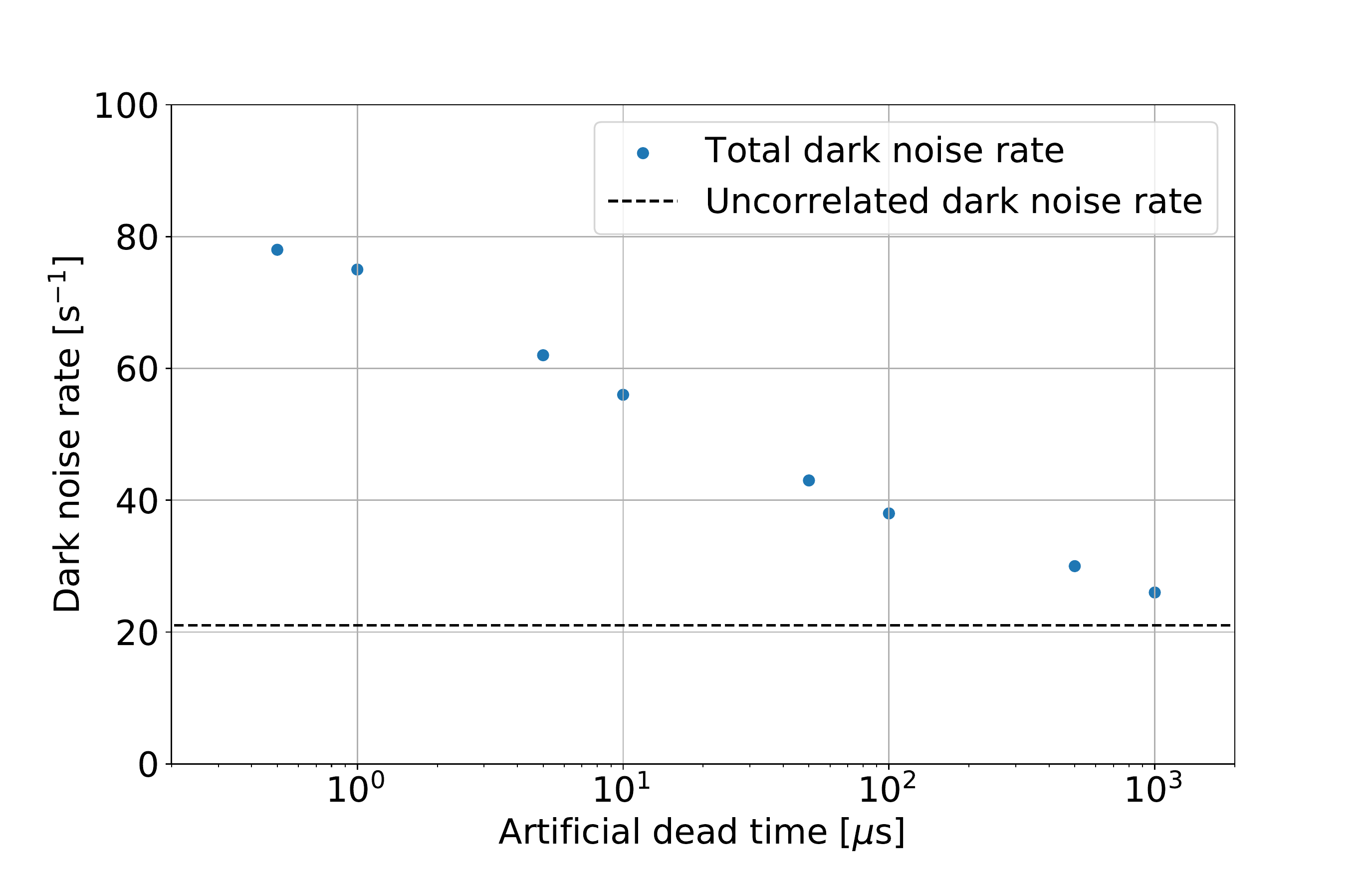}
        \caption{Effect of the artificial dead time window on the total dark noise for the 3.5 inch XP82B20D PMT with serial number SN80171 at a temperature of -35\,$^{\circ}$C. As the dead time window increases from 0.5 to 1000 microseconds, the total dark noise rate approaches the uncorrelated dark noise rate (horizontal dashed black line) by cutting out correlated subsequent pulses at small time differences.}
        \label{fig:deadtimewindow}
\end{figure}

The contribution from correlated dark noise can be reduced by introducing the concept of an artificial dead time. By ignoring all subsequent hits within a certain time window (reducing the total number of events), the correlated events are gradually filtered out with increasing dead time window. This is illustrated in figure~\ref{fig:deadtimewindow}, where the total dark noise rate for higher dead time windows starts approaching the uncorrelated dark noise rate.

\subsection{Dark Noise as a Function of Temperature}

\begin{figure}[ht]
\centering
  \includegraphics[width=0.9\textwidth]{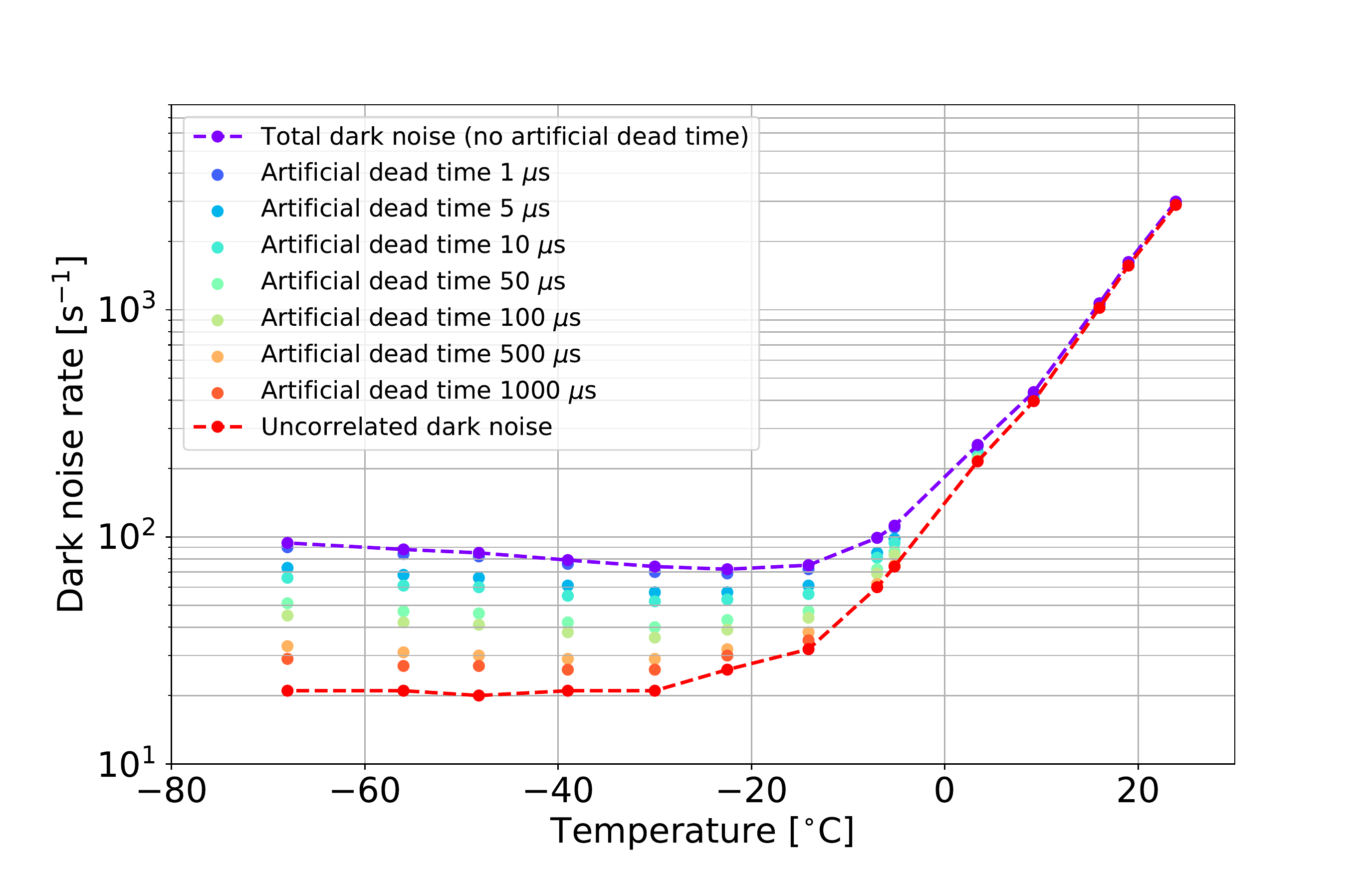} 
\caption{Dark noise rates for various artificial dead time windows and fitted uncorrelated dark noise rates as a function of temperature for the 3.5 inch XP82B20D PMT with serial number SN80169.}
 \label{fig:darknoisefigure}
\end{figure}

For all PMTs, both the total dark noise rate as a function of increasing dead time window and the uncorrelated dark noise rate are measured as a function of temperature. The chosen temperature range is -70\,$^{\circ}$C to 25\,$^{\circ}$C. As an example, these measurements are shown for one particular PMT in figure~\ref{fig:darknoisefigure}. For all other PMTs, dark noise rates at two benchmark temperatures (-30\,$^{\circ}$C and 20\,$^{\circ}$C) are summarised in table~\ref{tab:summarysmall} and table~\ref{tab:summarylarge} for the 3.5 inch XP82B20D PMTs and the 9 inch XP1805D PMTs respectively, including a comparison with HZC dark noise rate data obtained at a temperature of 20\,$^{\circ}$C.

\subsection{Dark Noise of Tape-Covered PMTs}
In order to further study the low-temperature dark noise rates of the PMTs, the photocathode area of several PMTs are covered with black tape. Good optical contact (i.e. no air bubbles trapped between the tape and the photocathode) ensures that photons that are liberated in the PMT glass \cite{Rawlins} are absorbed instead of possibly being internally reflected towards the photocathode. In that way, the correlated noise (and therefore the total noise) is expected to decrease.

The rationale for these taping tests is provided by in-situ measurements in IceCube optical modules, in which case there is good optical coupling between the PMT glass, the optical module glass and the surrounding ice, resulting in low dark noise rates \cite{IceCubePMT}. In the lab, the optical coupling between a bare PMT and the surrounding air is not as good and the tape provides a way to mimic the in-ice conditions better. 

\begin{figure}[ht]
    \centering
        \includegraphics[width=0.4\textwidth]{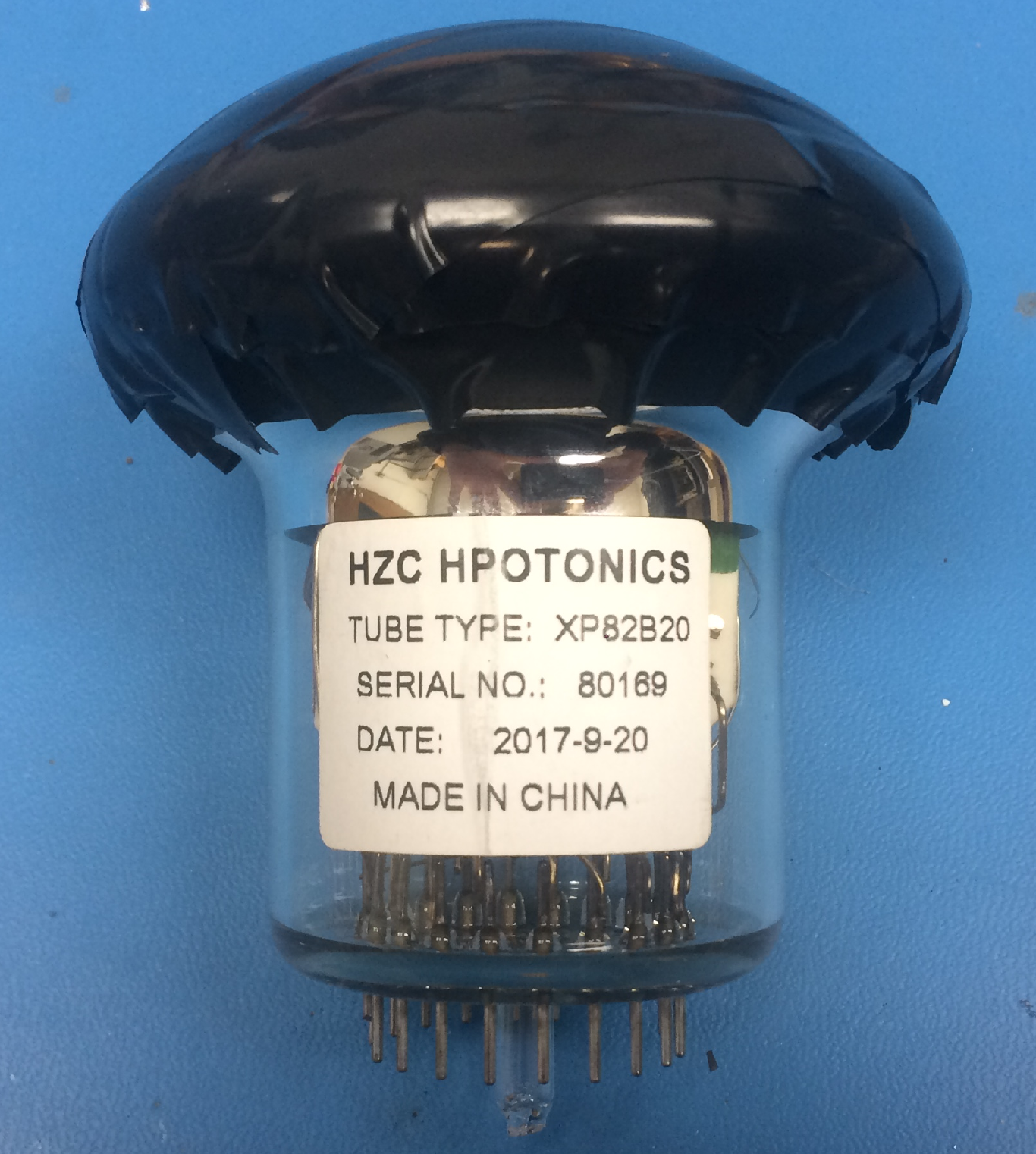}
        \caption{Picture of the 3.5 inch XP82B20D PMT with serial number SN80169 covered in black tape to absorb photons released in the PMT glass during dark noise measurements.}
        \label{fig:coveredpmt}
\end{figure}

\begin{figure}[ht]
\centering
\includegraphics[width=0.9\textwidth]{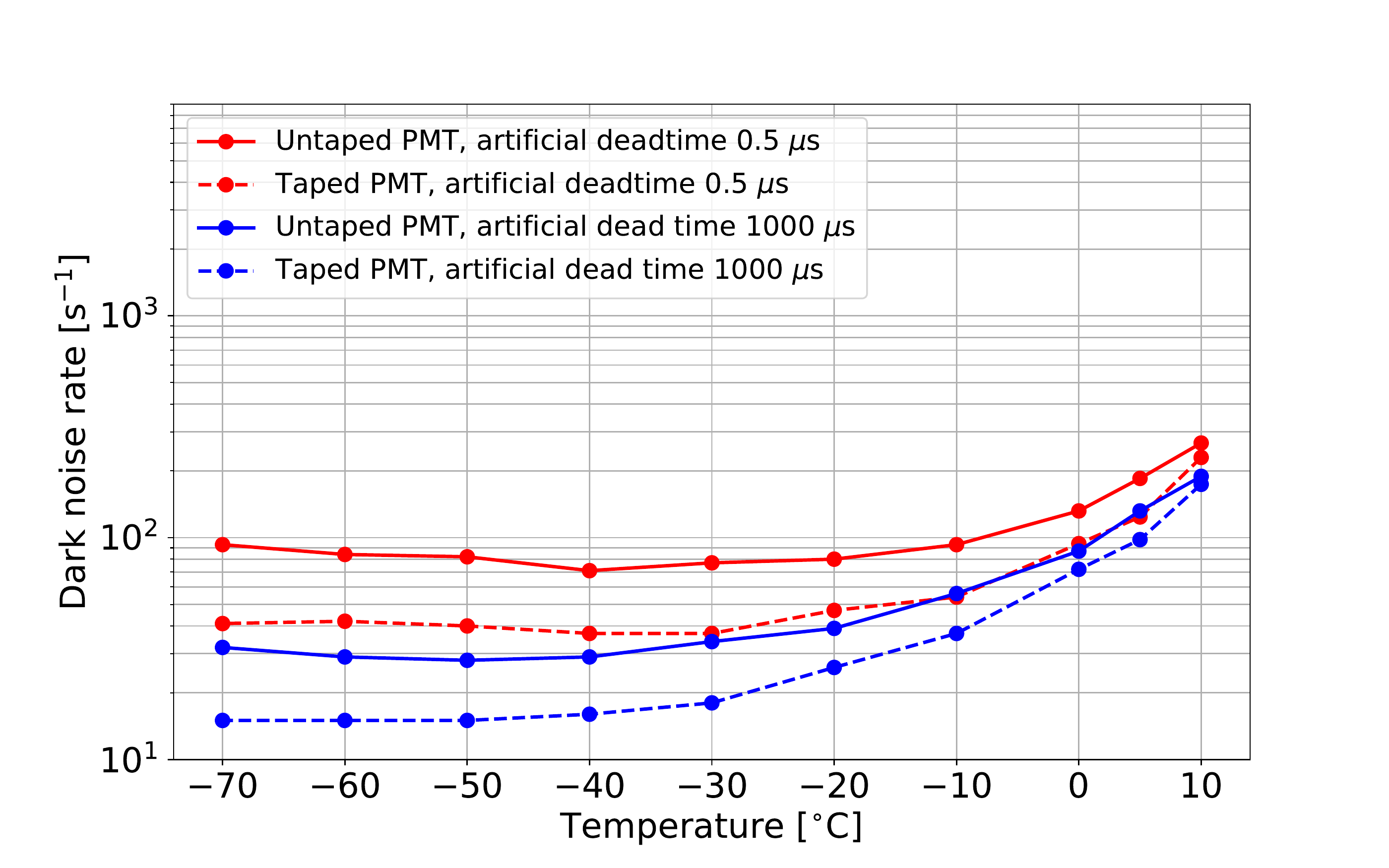} 
\caption{Dark noise rate as a function of temperature for the 3.5 inch XP82B20D PMT with serial number SN80169 at an artificial dead time of 0.5\,$\mu$s (red) and 1000\,$\mu$s (blue), comparing the bare untaped PMT (solid lines) and the PMT covered by black tape (dashed lines) to absorb photons released in the PMT glass during dark noise measurements. For the taped PMT, dark noise rates are further reduced by more than a factor of two at low temperatures compared to the untaped PMT.}
\label{fig:tapedvsuntaped}
\end{figure}

A picture of a tape-covered PMT is shown in figure~\ref{fig:coveredpmt}. Figure~\ref{fig:tapedvsuntaped} shows a comparison between the dark noise rates as a function of temperature for an untaped and a tape-covered PMT at an artificial dead time of 0.5\,$\mu$s and 1000\,$\mu$s. At low temperatures, the observed reduction in dark noise rate due to the tape is roughly a factor of two. This can be explained by the following back-of-the-envelope calculation: the critical angle for total internal refraction between borosilicate glass ($n=1.52$) and air ($n=1.00$) is 41 degrees. Assuming that photons are liberated in the glass at uniformly distributed angles, the fraction of photons escaping the glass (and thus detection) is the solid angle of the critical angle cone divided by 4$\pi$ which is equal to 0.12. When the PMT is covered in tape, there is no glass-air interface and half of the photons escape detection due to absorption by the tape. The expected reduction in number of photons reaching the photocathode is thus $\frac{1-0.12}{0.5}=1.8$, which is close to the observed number.

\section{Conclusion}\label{sec:conclusion}
The IceCube Collaboration is investigating various types of PMTs for possible use in future optical modules. This report summarises PMT characterization studies for two types of HZC PMTs: the 3.5 inch XP82B20D and the 9 inch XP1805D. The main results are summarised in table~\ref{tab:summarysmall} and table~\ref{tab:summarylarge} for the 3.5 inch XP82B20D PMTs and the 9 inch XP1805D PMTs, respectively and are in good agreement with the specifications as provided by HZC. In addition, excellent noise behaviour is observed at the low temperatures relevant for possible use in future IceCube optical modules.

\begin{table}[ht]
\begin{centering}
\footnotesize
\begin{tabular}{|p{4.5cm}llll|}
\hline
 & \textbf{SN80187} & \textbf{SN80171} & \textbf{SN80169} & \textbf{HZC Specification} \\
\hline
Gain slope (log/log) & $7.0 \pm 0.1$ & $6.6 \pm 0.1$ & $6.6 \pm 0.1$ & $6.5-8.0$ \\
Early pulses [\%] & $0.85 \pm 0.02$ & $0.77 \pm 0.02$ & $0.79 \pm 0.02$ & $1-1.5$ \\
Late pulses [\%] & $4.00 \pm 0.05$ & $4.22 \pm 0.06$ & $3.35 \pm 0.04$ & $2.5-5.5$ \\
Transit time spread [ns] & $3.71 \pm 0.05$ & $4.15 \pm 0.16$ & $2.93 \pm 0.04$ & $2-4$ \\
Total DNR [$\mathrm{{s}}^{{-1}}$] at 20$^{\circ}$C & $339 \pm 1 \,(217)$ & $449 \pm 1 \,(667) $ & $1623 \pm 4 \,(876) $ & $800-2000$ \\
Uncorrelated DNR [$\mathrm{{s}}^{{-1}}$] at 20$^{\circ}$C& $306 \pm 1$ & $415 \pm 2$ & $1565 \pm 6$ & $800-2000$ \\
Total DNR [$\mathrm{{s}}^{{-1}}$] at -30$^{\circ}$C & $65.9 \pm 0.2$ & $71.7 \pm 0.2$ & $73.9 \pm 0.2$ & $-$ \\
Uncorrelated DNR [$\mathrm{{s}}^{{-1}}$] at -30$^{\circ}$C & $17.5 \pm 0.1$ & $20.7 \pm 0.2$ & $21.2 \pm 0.2$ & $-$ \\
\hline
\end{tabular}
\caption{Table summarising all 3.5 inch XP82B20D PMT measurements, compared to the HZC specifications (if available) in the last column. Transit time spread is the full-width half-maximum (FWHM) of a Gaussian fit to the relative transit time distribution. For the total dark noise rate (DNR) at 20\,$^{\circ}$C, the number in parentheses indicates the measurement performed by HZC as reported on the PMT datasheet.}
\label{tab:summarysmall}
\end{centering}
\end{table}

\begin{table}[ht]
\begin{centering}
\footnotesize
\begin{tabular}{|llll|}
\hline
& \textbf{SN00492} & \textbf{SN00386} & \textbf{HZC Specification} \\
\hline
Gain slope (log/log) & $5.1 \pm 0.1$ & $5.1 \pm 0.1$ & $5.5$ \\
Early pulses [\%] & $-$ & $-$  & $-$ \\
Late pulses [\%] & $5.17 \pm 0.06$ & $3.61 \pm 0.05$ & $-$ \\
Transit time spread [ns] & $2.89 \pm 0.07$ & $2.97 \pm 0.08$  & $2.4$ \\
Total DNR [$\mathrm{{s}}^{{-1}}$] at 20$^{\circ}$C & $6020\pm 20 \,(4665)$ & $901 \pm 2 \,(1376)$  & $3000-5000$ \\
Uncorrelated DNR [$\mathrm{{s}}^{{-1}}$] at 20$^{\circ}$C & $5870 \pm 20$ & $745 \pm 4$  & $3000-5000$ \\
Total DNR [$\mathrm{{s}}^{{-1}}$] at -30$^{\circ}$C & $488 \pm 1$ & $421 \pm 1$  & $-$ \\
Uncorrelated DNR [$\mathrm{{s}}^{{-1}}$] at -30$^{\circ}$C & $156 \pm 2$ & $145 \pm 1$  & $-$ \\
\hline
\end{tabular}
\caption{Table summarising all 9 inch XP1805D PMT measurements, compared to the HZC specifications (if available) in the last column. Transit time spread is the full-width half-maximum (FWHM) of a Gaussian fit to the relative transit time distribution. For the total dark noise rate (DNR) at 20\,$^{\circ}$C, the number in parentheses indicates the measurement performed by HZC as reported on the PMT datasheet.}
\label{tab:summarylarge}
\end{centering}
\end{table}


\acknowledgments
D.\,van Eijk would like to thank the International Balzan Prize Foundation and Francis Halzen for their support.



\end{document}